\documentclass[draft]{agujournal2019}
\usepackage[english]{babel}
\usepackage{amsmath}
\usepackage{txfonts}
\usepackage{lineno}
\usepackage{setspace}
\usepackage{multirow}
\usepackage{color}
\journalname{JGR: Planets}
\begin{document}
\title{Application of the mixing length theory to assess the generation of melt in internally heated systems}
\authors{K. Vilella \affil{1,2} and S. Kamata\affil{2}}
\affiliation{1}{JSPS International Research Fellow, Hokkaido University, Japan}
\affiliation{2}{Department of Earth and Planetary Sciences, Faculty of Science, Hokkaido University, Japan}
\correspondingauthor{Kenny Vilella}{kennyvilella@gmail.com}
\begin{keypoints}
\item We develop a new 1D analytical approach based on an extension of the mixing length theory.
\item Our approach is able to estimate the evolution of a system including the effects of melting, secular cooling, and radioactive heating.
\item We verify the validity of our approach by comparing its prediction with results obtained in 3D numerical simulations.
\end{keypoints}
\begin{abstract}
The effect of melting in planetary mantles plays a key role in their thermo-chemical evolution. 
Because of the laterally heterogeneous nature of melting, 3D numerical simulations
are in principle necessary prohibiting us from exploring wide ranges of conditions.
To overcome this issue, we propose a new analytical framework allowing to estimate the amount and depths of melting in a 1D analytical model for a simplified convective system.
To do so, we develop an approach, partly based on an extended version of the mixing length theory, able to estimate the distribution of the hottest temperatures in natural systems.
The approach involves several free parameters that are calibrated by fitting 3D numerical simulations.
We demonstrate that our algorithm provides the melting profile at steady-state as well as the
long-term evolution in fairly good agreement with the ones obtained in 3D numerical simulations.
We then apply our framework for a wide variety of planetary sizes and heating rates.
We find that an increase in planetary size increases the depth of melting for small planets but decreases for large planets.
This change in the trend is caused by the pressure dependence of the solidus.

\end{abstract}
\section*{Plain Language Summary}
The thermal evolution of terrestrial planets is traditionally investigated using 3D numerical simulations.
However, terrestrial planets are often poorly constrained making a thorough investigation very time-consuming.
One way to cope with this issue is simply to use much faster models, such as analytical models, i.e., models that only estimate the 1D evolution.
One drawback of this approach is that melting and the subsequent volcanism has to be neglected, while it is an important feature of planetary evolution.
Here, we build on previous analytical models to propose a new framework allowing to estimate the amount and depths of melting in a 1D analytical model.
The accuracy of our framework is verified by comparing its prediction with results obtained in 3D numerical simulations.
In particular, when applied to a generic planetary mantle, our algorithm provides a long-term evolution in fairly good agreement with the one obtained in 3D numerical simulations.
At last, we use our approach to assess the generation of melt in terrestrial planets with various radius and amount of radioactive isotopes.
\section{Introduction \label{secIntro}}

The thermo-chemical evolution of a planetary body is fundamental to understand its past and current state.
For a terrestrial planet, this evolution is characterized by the coupling of different processes occurring in different parts of the planet and at different temporal and spatial scales.
An usual way to cope with the difficulty of modelling such a complex system is to focus on the long term evolution of the planetary mantle.
Indeed, because of its much slower dynamics compared to the atmosphere or the liquid iron core, the evolution of the mantle is believed to control the heat transfer of the whole planet.
However, dynamics of planetary mantles is in itself a very complex problem, as one may face several major technical difficulties before obtaining an accurate modelling.

One major issue is to account for the melting of rocks and the subsequent chemical differentiation and volcanism.
Although increasingly sophisticated modelling of melting is considered in 3D numerical simulations \cite{Christensen1994, Samuel2003, Xie2004, Lourenco2018}, the composition of the produced melt is generally prescribed and is independent of the depth and temperature for which melting occurred.
While this simplification is required in the light of the amount of computational resources available, it induces bias in the estimated long-term evolution, especially the location, quantity and composition of melt.

Alternatively, we propose to account for the effects of melting on planets evolution using a 1D analytical model \cite<e.g.,>{Kamata2018}.
Such an approach has the important advantage to be computationally inexpensive allowing to investigate problems that are out of reach of 3D numerical simulations.
In particular, the mixing length theory provides a robust tool to estimate, at first order, the laterally averaged temperature profile of a system \cite{Vitense1953, Spiegel1963, Sasaki1986}.
This method has been used to study the thermal evolution of magma ocean \cite{Abe1997}, icy satellites \cite{Kamata2018}, or terrestrial planets \cite{Wagner2019}.
However, the modelling of melting requires not only the estimate of the average temperature profile, but also that of the full lateral distribution of temperature.
Indeed, using the average temperature profile to account for melting will lead to a large underestimation of the amount of melting and an overestimation of the melting depths \cite<see for instance the temperature distributions reported in>[]{Vilella2018b}.
As a matter of fact, considering the typical convective structure of a terrestrial planet, an average temperature profile that is locally slightly larger than the solidus temperature would actually indicate the presence of a local magma ocean.
Therefore, except for the peculiar case of Io \cite{Steinke2020}, melting is traditionally not included in parametrized convection or in 1D analytical calculation.

In this work, we derive a new analytical framework allowing to estimate the amount and depths of melting in 1D evolution model.
As 1D models are not appropriate for estimating the full lateral distribution of temperature, we rather develop an approach, partly based on an extended version of the mixing length theory, able to estimate the cumulative distribution function of the hottest temperatures.
In other words, our aim is not to constrain the precise lateral distribution of temperature but only the distribution of the hottest temperatures from a statistic/probability point of view.
Due to the novelty of our approach, we apply our framework to a simplified convective system consisting of a purely internally heated fluid in a Cartesian geometry.
This approach involves several free parameters that are calibrated by fitting 3D numerical simulations \cite<obtained previously in>[]{Limare2015, Vilella2018b}.
The reliability of our analytical approach is then evaluated by comparing the estimated amount of melting with the one predicted by 3D numerical simulations.
In that aim, we develop an algorithm allowing to calculate the evolution of a system experiencing secular cooling and melting.
At last, this algorithm is used to investigate the influence of different parameters on the amount and depths of melting.

\section{Theoretical considerations \label{secTheoretical}}

Before developing our extended version of the mixing length theory (MLT), we give a short description of the convective system considered and a brief overview of the classical MLT.
These preliminary considerations are crucial to understand the limitations and strengths of our theoretical approach.

\subsection{Description of purely internally heated convection \label{secDescription}}

As a reference set-up for our approach, we consider a purely internally heated fluid.
This convective system generally consists in a horizontal layer of fluid, with a constant temperature imposed at the top boundary and an adiabatic condition at the base.
Because of the adiabatic condition at the bottom, the only source of heat is volumetric heating, i.e., each parcel of fluid is homogeneously heated, and the heat escapes from the top boundary.
When the fluid is further considered to be isoviscous and incompressible, the system is only controlled by the Rayleigh-Roberts number \cite{Roberts1967},
\begin{linenomath*} 
\begin{equation}
Ra_{H} = \frac{\rho g \alpha \Delta T_{H} d^{3}}{\kappa \eta},
\label{eqRaH}
\end{equation}
\end{linenomath*}
where 
\begin{linenomath*} 
\begin{equation}
\Delta T_{H} = \frac{H d^{2}}{\lambda}
\label{eqDTH}
\end{equation}
\end{linenomath*}
is the temperature scale of the system, $\rho$ density, $g$ the acceleration of gravity, $\alpha$ the thermal expansion coefficient, $H$ the volumetric heating rate, $d$ the layer's height, $\lambda$ the thermal conductivity, $\kappa$ the thermal diffusivity, and $\eta$ viscosity.
The Rayleigh-Roberts number quantifies the vigour of convection.
With increasing $Ra_{H}$, the convective system becomes more and more chaotic, velocities are increasing and the lengthscale of convective structure is decreasing.
As this convective system has already been extensively investigated \cite{Parmentier2000, Vilella2017}, we will only mention few of its important characteristics.
Note that, here and hereafter, we will use for illustration purposes the 3D numerical simulations published in \citeA{Vilella2018b} and obtained with StagYY \cite{Tackley2008}.

\begin{figure}[t!]
\begin{center}
\includegraphics[width=0.90\textwidth]{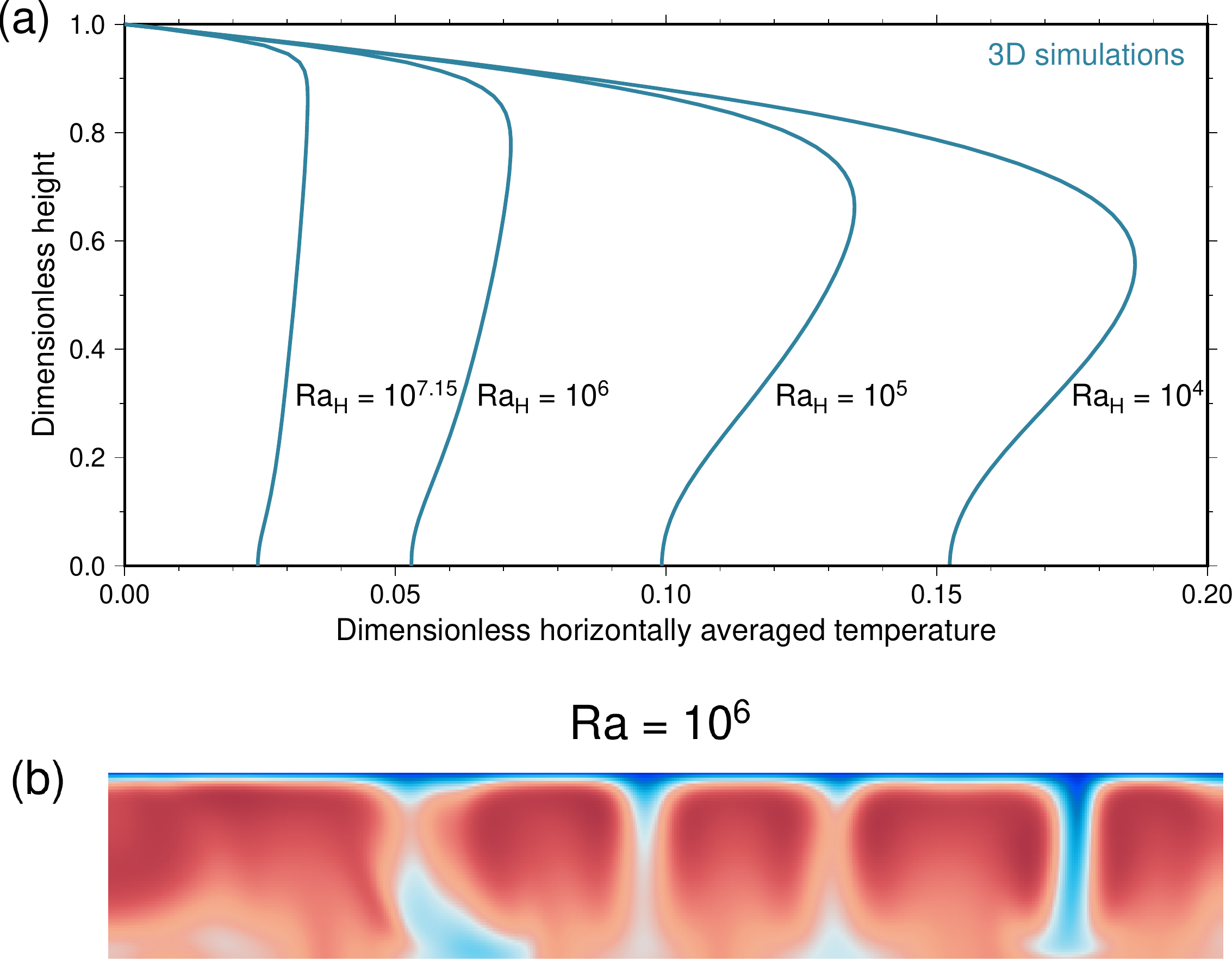}
\end{center}
\caption{\label{TProfil} (a) Horizontally averaged temperature profiles obtained for various values of the Rayleigh-Roberts number ($Ra_{H}$). (b) Vertical slice of the temperature field obtained for $Ra_{H} = 10^{6}$. Blue and red colours correspond to cold and hot temperatures, respectively. The results of (a) and (b) are from 3D numerical simulations \cite<see>[for more details on the simulations]{Vilella2018b}.}
\end{figure}

A brief look at the convective system (figure~\ref{TProfil}) indicates the presence of a top thermal boundary layer (TBL), where large vertical temperature variations occur, above an almost isothermal convective interior.
The top TBL controls the dynamics of the system by generating cold downwellings that sink into the convective interior.
As a result, a non-buoyant return flow is present in the convective interior to balance the input of sinking materials.
However, a part of the cold material is accumulating at the base of the system (figure~\ref{TProfil}b) inducing a slight decrease of the temperature with depth (figure~\ref{TProfil}a).
This feature is called subadiabaticity and has been observed in various internally heated systems \cite<e.g,>[]{Jeanloz1987, Sinha2007}.

\subsection{Fundamentals of the mixing length theory}

The 1D thermal evolution of a system is mainly governed by its conservation of energy, which at steady state can be written as
\begin{linenomath*} 
\begin{equation}
0 = -\frac{\mathrm{d}F_{cond}}{\mathrm{d}z} - \frac{\mathrm{d}F_{conv}}{\mathrm{d}z} + H,
\label{eqEnergy}
\end{equation}
\end{linenomath*}
where $z$ is height, $F_{cond} = - \lambda \mathrm{d}T / \mathrm{d}z$ the conductive heat flux, and $F_{conv}$ the advective heat flux.
The further integration of eq.~\ref{eqEnergy} gives simply,
\begin{linenomath*} 
\begin{equation}
\text{C} = -F_{cond} - F_{conv} + Hz,
\label{eqMLTdim}
\end{equation}
\end{linenomath*}
with $\text{C}$ a constant to be determined.
At the base of the system ($z = 0$), $F_{cond} = 0$ and $F_{conv} = 0$, because the base is adiabatic, implying $\text{C} = 0$.
As a consequence, estimating the thermal evolution of the system consists in the determination of the advective heat flux $F_{conv}$.
The purpose of the MLT is to obtain an approximate expression of this advective heat flux by neglecting the horizontal advection of heat, so that $F_{conv}$ simply represents the vertical advection of heat.
In that purpose, one can write, at first order,
\begin{linenomath*} 
\begin{equation}
F_{conv} = \overline{\rho C_{p} w \theta},
\label{eqConv}
\end{equation}
\end{linenomath*}
where $C_{p} = \lambda / \alpha \rho$ is the specific heat capacity, $w$ the vertical velocity, and $\theta$ the temperature perturbation.
Note that here and hereafter the overbar denotes average properties.
Assuming a Stokes velocity for $w$, determining $F_{conv}$ boils down in the determination of an appropriate expression for $\theta$.
To do so, the MLT focuses on a column of fluid and considers that the perturbation of temperature $\theta$ is caused by a fluid parcel transported from a height $z = z_{0}$ that was originally at ambient temperature $\overline{T(z_{0})}$.
This implies
\begin{linenomath*} 
\begin{equation}
T(z) = \overline{T(z_{0})} + \frac{\mathrm{d}T}{\mathrm{d}z}(z - z_{0}).
\end{equation}
\end{linenomath*}  
Furthermore, by a lateral average of this equation we obtain
\begin{linenomath*} 
\begin{equation}
\overline{T(z)} = \overline{T(z_{0})} + \frac{\mathrm{d}\overline{T}}{\mathrm{d}z}(z - z_{0}),
\end{equation}
\end{linenomath*}  
and when combining these two equations,
\begin{linenomath*} 
\begin{equation}
\theta = T(z) - \overline{T(z)} = \biggl(\frac{\mathrm{d}T}{\mathrm{d}z} - \frac{\mathrm{d}\overline{T}}{\mathrm{d}z}\biggr)(z - z_{0}).
\label{eqTheta}
\end{equation}
\end{linenomath*}  
We further note $l = 2 (z - z_{0})$ the mixing length, i.e., corresponding to the typical length-scale where temperature perturbations are homogenized.
Finally, we incorporate the Stokes velocity and eq.~\ref{eqTheta} into eq.~\ref{eqConv} to obtain
\begin{linenomath*} 
\begin{equation}
F_{conv} = \frac{\rho^{2}C_{p}\alpha g l(z)^{4}}{18\eta}\biggl(\frac{\mathrm{d}T}{\mathrm{d}z} - \frac{\mathrm{d}\overline{T}}{\mathrm{d}z}\biggr)^{2}.
\label{eqFconv}
\end{equation}
\end{linenomath*}
Incorporating eq.~\ref{eqFconv} into eq.~\ref{eqMLTdim} and removing the dimension from the equation, we finally obtain
\begin{linenomath*} 
\begin{equation}
\frac{\mathrm{d}T^{*}}{\mathrm{d}z^{*}} - \frac{Ra_{H} l^{*}(z^{*})^{4}}{18}\biggl(\frac{\mathrm{d}T^{*}}{\mathrm{d}z^{*}} - \frac{\mathrm{d}\overline{T^{*}}}{\mathrm{d}z^{*}}\biggr)^{2} + z^{*} = 0,
\label{eqMLT}
\end{equation}
\end{linenomath*}
where dimensionless values are denoted with a $^{*}$.
Equation~\ref{eqMLT} is the main equation solved in the following section.

\section{Extended version of the mixing length theory \label{secMLT}}

Our goal is to build an analytical framework allowing to estimate the distribution of the hottest temperatures.
The achievement of this goal will be carried out in several stages.
First, we propose a modification of the mixing length theory providing a satisfactory estimate of the horizontally averaged temperature profile for purely internally heated fluids.
We then apply this framework to determine the ``hot'' temperature profile, i.e., composed of the hottest temperature at every depth.
Finally, using the estimated hot and average temperature profiles, we develop a process to reconstruct the cumulative distribution function for the 5\% hottest temperatures.

\subsection{Determination of the average temperature profile $T^{*}_{avg}$}
\begin{figure}[t!]
\begin{center}
\includegraphics[width=0.9\textwidth]{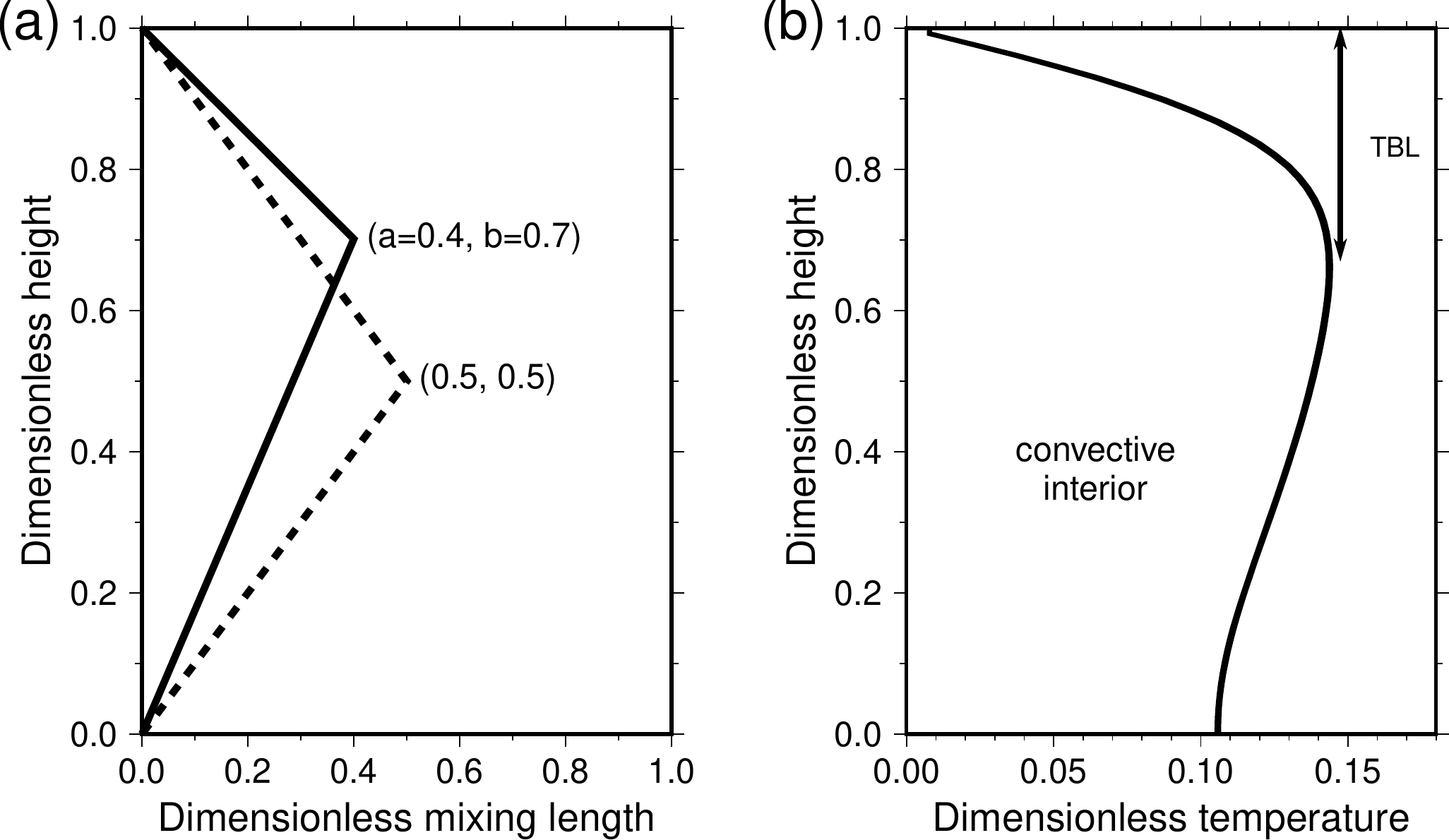}
\end{center}
\caption{\label{MLT_Kamata} (a) Plots of the dimensionless mixing length ($l^{*}$) and (b) a horizontally averaged temperature profile as a function of the dimensionless height. For the sake of simplicity, we characterize $l^{*}$ with (a,b) where $l^{*}(b) = a$ is the maximum value of $l^{*}$. The dashed line represents the traditional mixing length (0.5, 0.5), while the solid line is an example of mixing length used here. Note that in our framework $a$ is given by eq.~\ref{eqLB} while $b = 1 - \delta_{TBL}$, where $\delta_{TBL}$ is the thickness of the thermal boundary layer.}
\end{figure}

The determination of the temperature profile using eq.~\ref{eqMLT} requires to prescribe a profile for $l^{*}$ and for $\mathrm{d}\overline{T^{*}} / \mathrm{d}z^{*}$.
While $l^{*}$ is generally assumed to be the closest distance from a horizontal boundary, the average temperature profile is difficult to evaluate \textit{a priori}. 
As such, the MLT is generally used to study turbulent convection, because, for this regime, the average temperature is simply the adiabatic temperature profile.
Although not strictly valid, \citeA{Kamata2018} has also used the adiabatic profile as a proxy for the average temperature profile in laminar convection. 
Provided that a modified expression for $l^{*}$ is considered (figure~\ref{MLT_Kamata}), this approach is able to reproduce reasonably well numerical results.

As a first step, we apply this formalism to internally heated convection. 
In that case, the adiabatic temperature profile is constant so eq.~\ref{eqMLT} simply writes
\begin{linenomath*} 
\begin{equation}
\frac{\mathrm{d}T^{*}}{\mathrm{d}z^{*}} - \frac{Ra_{H} l^{*}(z^{*})^{4}}{18}\biggl(\frac{\mathrm{d}T^{*}}{\mathrm{d}z^{*}}\biggr)^{2} + z^{*} = 0.
\label{eqMLT_moy}
\end{equation}
\end{linenomath*}
As one may have noted, this equation does not have solution for $\mathrm{d}T^{*} / \mathrm{d}z^{*} = 0$ and $z^{*} > 0$.
This means that the temperature will continuously increase or decrease with depth.
In practice, because the average temperature profile is maximum at the base of the TBL (figure~\ref{TProfil}), only the upper part of the TBL can be reproduced correctly.
This formalism is therefore not appropriate for the specific case of purely internally heated systems.

Alternatively, we propose to approximate $\mathrm{d}\overline{T^{*}}/ \mathrm{d}z^{*}$ using a simple but unknown function ($f^{*}$) of $z$.
By doing this, we depart from the framework used to establish eq.~\ref{eqMLT}, but it has the advantage of giving solutions for purely internally heated fluids.
For the specific case $\mathrm{d}T^{*} / \mathrm{d}z^{*} = 0$, obtained at $z^{*} = 1-\delta_{TBL}^{*}$, eq.~\ref{eqMLT} implies
\begin{linenomath*} 
\begin{equation}
l^{*}(z = 1 - \delta_{TBL}^{*}) = \biggl(\dfrac{18(1 - \delta_{TBL}^{*})}{Ra_{H}f^{*}(z = 1 - \delta_{TBL}^{*})^{2}} \biggr)^{1/4},
\label{eqLB}
\end{equation}
\end{linenomath*}
with $\delta_{TBL}^{*}$ the thickness of the TBL.
We can further assume that the mixing length is maximum at $z^{*} = 1-\delta_{TBL}^{*}$ (figure~\ref{MLT_Kamata}).
In that case, solving eq.~\ref{eqMLT} only requires to determine $f^{*}$ and $\delta_{TBL}^{*}$, the latter can be obtained from numerical simulations, while the former has to be searched manually.
Therefore, the main difficulty of our approach lies in the determination of a robust and convenient expression for $f^{*}$.

There are several possible methods to achieve this task.
For instance, one may conduct a systematic search using a large range of function to find one providing a good fit of the whole temperature profile.
The risk of this method is to require a modelling of $f^{*}$ so complex that it decreases the applicability of our approach.
As a consequence, our emphasis is rather on finding a simple expression, i.e., without an excessive number of parameters, that provides an accurate description of the thermal boundary layer.
In other words, since most melting is occurring close to the surface of terrestrial planets, we decided to favour the simplicity of the model over its ability to reproduce well the subadiabaticity.
As such, we simply consider
\begin{linenomath*} 
\begin{equation}
f^{*}(z^{*}) = c/z^{*\, d},
\label{eqMoy_f}
\end{equation}
\end{linenomath*}
where $c$ and $d$ are constants.

\begin{figure}[t!]
\begin{center}
\includegraphics[width=0.9\textwidth]{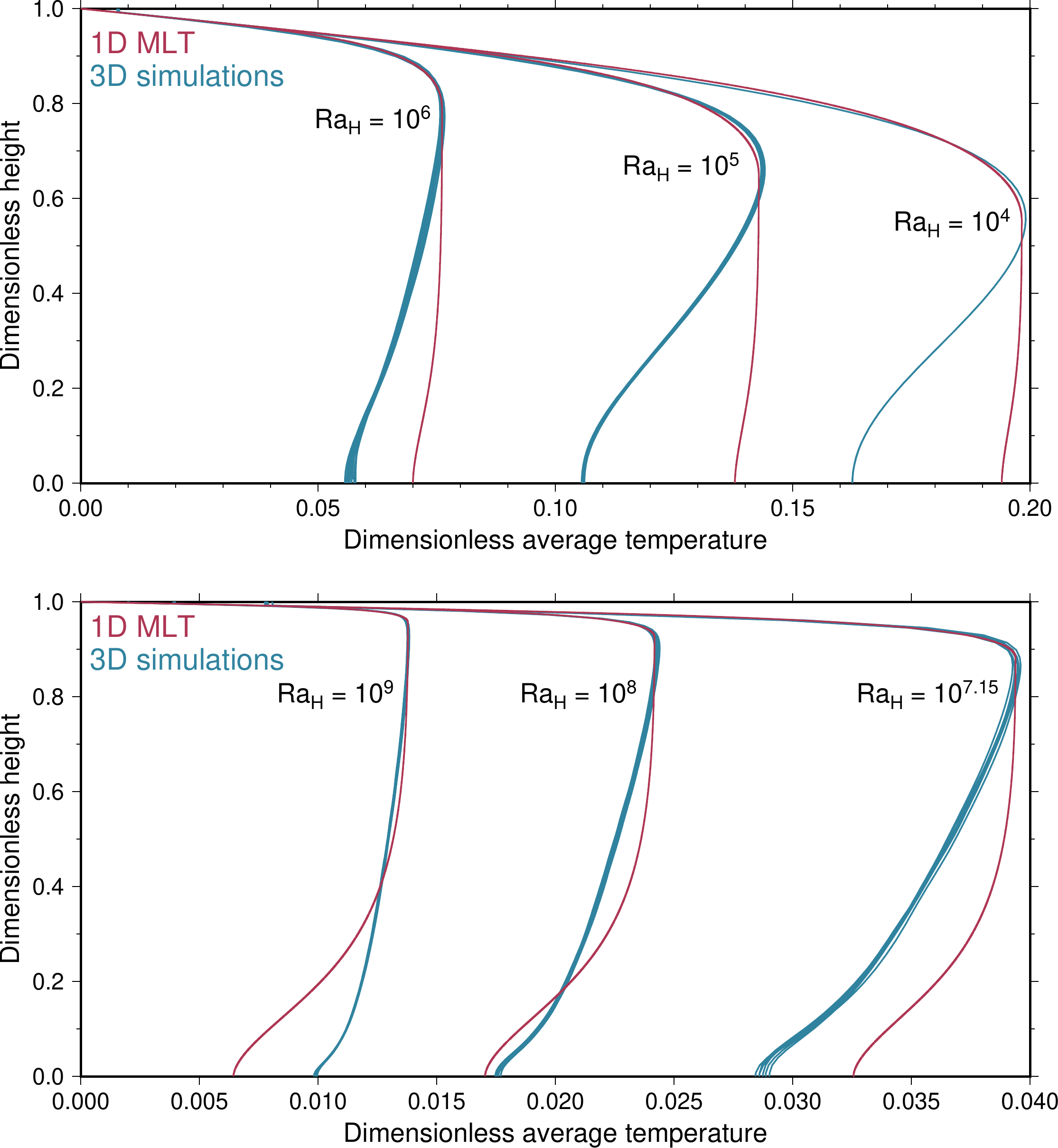}
\end{center}
\caption{\label{Profile_Tmoy} Horizontally averaged temperature profiles from 3D numerical simulations (blue curves) and calculated using the mixing length theory (red curves) for various values of the Rayleigh-Roberts number ($Ra_{H}$). Profiles at different time-steps are reported to account for temporal variations.}
\end{figure}

The next step is to verify the ability of our framework to reproduce numerical results.
To do so, we calculate the average temperature profile by solving eq.~\ref{eqMLT} (method described in Appendix) for a large range of parameter values ($c$, $d$ and $\delta_{TBL}^{*}$).
We then compare these analytical temperature profiles with our 3D numerical results.
The comparison reveals that analytical results are able to fit relatively well the temperature profile from each of our 3D numerical simulations.
In order to build a practical framework, it is now necessary to establish scaling relationships for the three parameters of the model.
Interestingly, our results indicate that it seems reasonable to consider a constant value for $c$ and $d$ implying a function $f^{*}$ that does not change with $Ra_{H}$ (values reported in Table~\ref{TabParam}).
$\delta_{TBL}^{*}$ is thus the only parameter function of $Ra_{H}$.
Previous works on similar convective systems \cite<e.g.,>[]{Sotin1999, Vilella2017} have shown that it is possible to establish a scaling relationship for $\delta_{TBL}^{*}$ using a power-law function.
Nevertheless, it is difficult to establish a single scaling relationship valid for the whole range of $Ra_{H}$ because of the change in the convective structure, e.g., from time-independent structure at $Ra_{H}<10^5$ to time-dependent structure at $Ra_{H}>10^5$.
Based on numerical results from \citeA{Vilella2017}, it seems more appropriate to separate the range of $Ra_{H}$ into three domains ($Ra_{H}<10^5$, $10^{5}\leq Ra_{H}<10^{7}$ and $Ra_{H} \geq 10^{7}$), and, for each domain, conduct a best-fit procedure using a power-law function.
The results of the best-fit procedure are reported in Table~\ref{TabParam}.
Using these values, we reconstruct the average temperature profile for each of our 3D numerical simulations.
Some representative results are shown in figure~\ref{Profile_Tmoy}.
The comparison shows a significant temperature deviation in the convective interior.  
Although striking, this deviation is not necessarily an issue for our purpose as the melt generation is expected to occur in the top TBL whose temperature profile is, by contrast, well reproduced.  

The modified version of the mixing length theory proposed in this section is therefore able to reproduce reasonably well the results of 3D numerical simulations, and this for all values of $Ra_{H}$ between $10^{4}$ and $10^{9}$.
The next step of our study is to apply this framework to the hot temperature profile.

\subsection{Determination of the hot temperature profile $T^{*}_{hot}$}
\begin{figure}[t!]
\begin{center}
\includegraphics[width=0.9\textwidth]{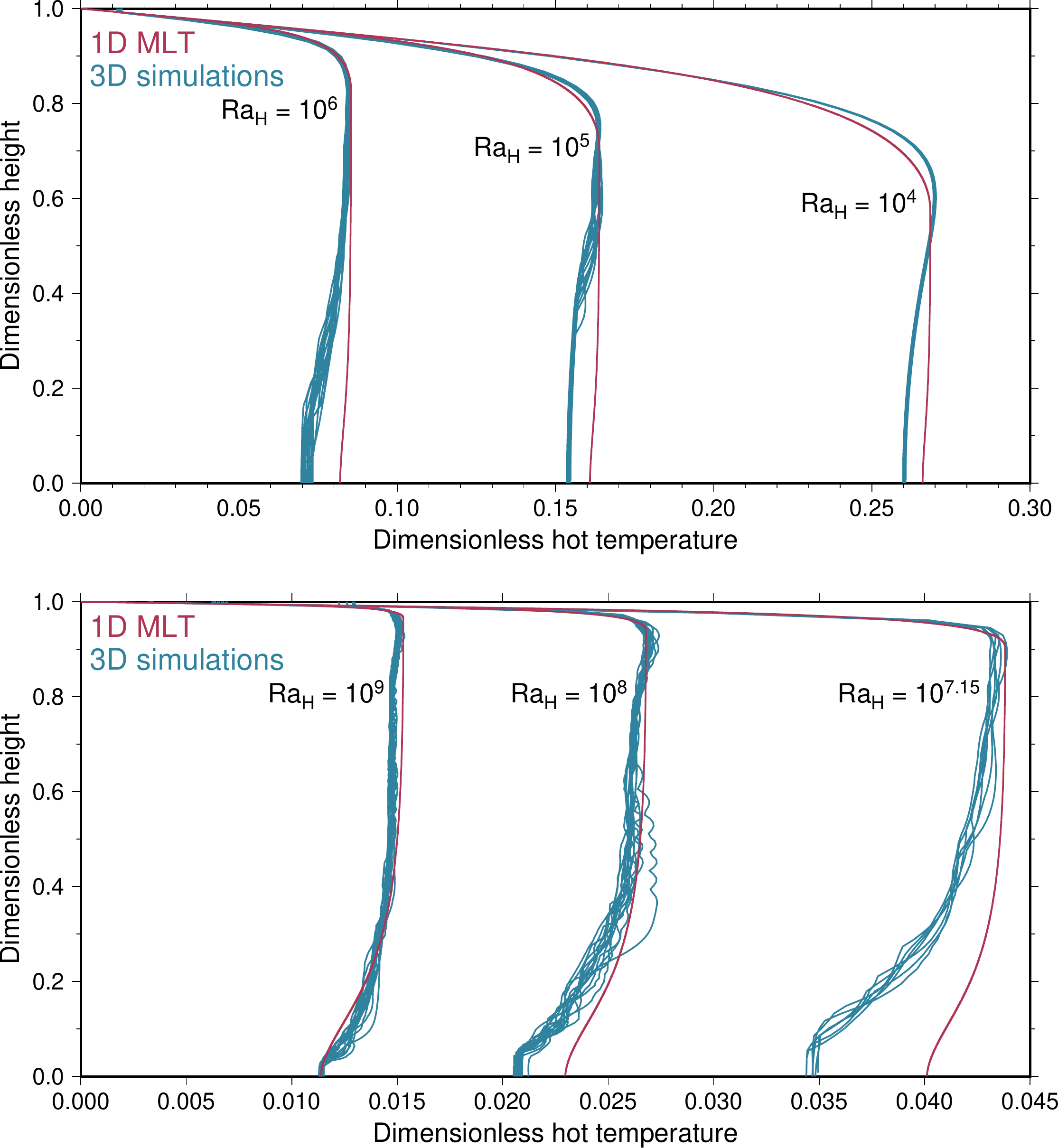}
\end{center}
\caption{\label{Profile_Tmax} Hot temperature profiles from 3D numerical simulations (blue curves) and calculated using the mixing length theory (red curves) for various values of the Rayleigh-Roberts number ($Ra_{H}$). Profiles at different time-steps are reported to account for temporal variations.}
\end{figure}

The hot temperature profile is essentially similar to the horizontally averaged temperature profile, with the exception of the surface heat flux.
Indeed, the average surface heat flux is, by construction, equal to $H d$, while the surface heat flux calculated from the hot temperature profile is a priori unknown.
Nevertheless, we found that its value is similar in all our numerical simulations, with a value between $1.52 H d$ and $1.75 H d$ (Supplementary Figure 1).
We therefore decide to assume an intermediate value of $1.65 H d$ and modify accordingly eq~\ref{eqMLT},
\begin{linenomath*} 
\begin{equation}
\frac{\mathrm{d}T^{*}}{\mathrm{d}z^{*}} - \frac{Ra_{H} l^{*}(z^{*})^{4}}{18}\biggl(\frac{\mathrm{d}T^{*}}{\mathrm{d}z^{*}} - f^{*}(z^{*})\biggr)^{2} + 1.65 z^{*} = 0.
\label{eqMLT_max}
\end{equation}
\end{linenomath*}
and eq~\ref{eqLB},
\begin{linenomath*} 
\begin{equation}
l^{*}(z = 1 - \delta_{TBL}^{*}) = \biggl(\dfrac{29.7(1 - \delta_{TBL}^{*})}{Ra_{H}f^{*}(z = 1 - \delta_{TBL}^{*})^{2}} \biggr)^{1/4}.
\label{eqLB_max}
\end{equation}
\end{linenomath*}
We can now estimate the model parameters ($c$, $d$ and $\delta_{TBL}^{*}$) for the hot temperature profile adopting the same procedure as for the average temperature profile.
The scaling relationship for each parameter is again reported in Table~\ref{TabParam}, while some representative results are shown in figure~\ref{Profile_Tmax}.
Compared to the average temperature profile (figure~\ref{Profile_Tmoy}), the fit of the TBL is slightly less satisfying, especially for low values of $Ra_{H}$.
By contrast, because the subadiabacity is less pronounced, the convective interior is better fitted.
Overall, the calculated hot temperature profiles are in good agreement with 3D numerical results.

\subsection{Determination of the temperature distribution}
\begin{figure}[t!]
\begin{center}
\includegraphics[width=0.9\textwidth]{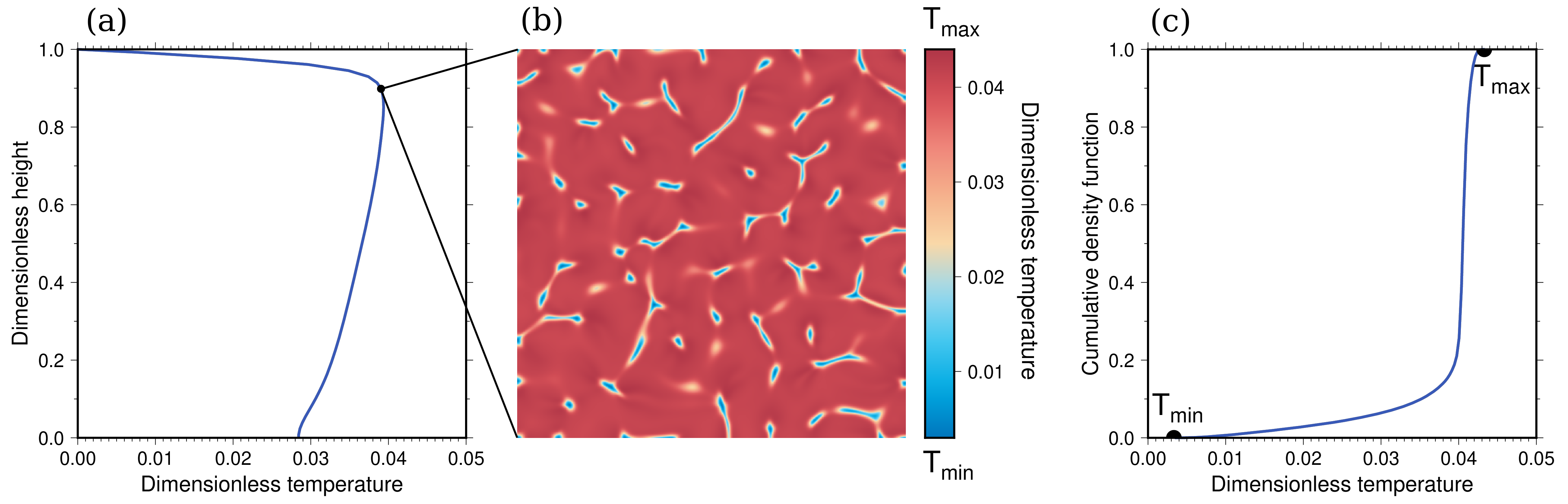}
\end{center}
\caption{\label{cdf} Illustration of the cumulative density function (cdf) using the 3D numerical simulation obtained for $Ra_{H} = 10^{7.15}$. At a given depth, here taken as a dimensionless height equal to 0.9, i.e., the base of the thermal boundary layer in (a) the average temperature profile, (b) the temperature field exhibit large lateral variations. These variations can be quantified using the (c) cdf. The cdf measures the proportion of material with a temperature lower than the temperature considered. For instance, a cdf of 0.5 indicates the median temperature, while a cdf of 0 and 1 indicate that there is not temperature lower or higher, respectively, than the temperature considered.}
\end{figure}

The last step is to estimate the distribution of the hottest temperatures using both the average and hot temperature profiles obtained above.
Here, we will mainly focus on the distribution of temperature at a given depth using the cumulative density function (cdf).
A brief explanation of the cdf is provided in figure~\ref{cdf}.
Although the precise shape of the distribution changes with $Ra_{H}$ \cite{Vilella2018b}, the main characteristics remain similar, independently of the $Ra_{H}$ and depth considered.
First, the cumulative density function slowly increases until values up to 0.1--0.2, then sharply increases until reaching 0.95--0.98, and again slowly increases until 1.0.
The fitting of this type of distribution is very challenging. 
In particular, the extreme values of the distribution will often be grossly misfitted by classical distribution functions, while it is for us the most important part.

\begin{figure}[t!]
\begin{center}
\includegraphics[width=0.99\textwidth]{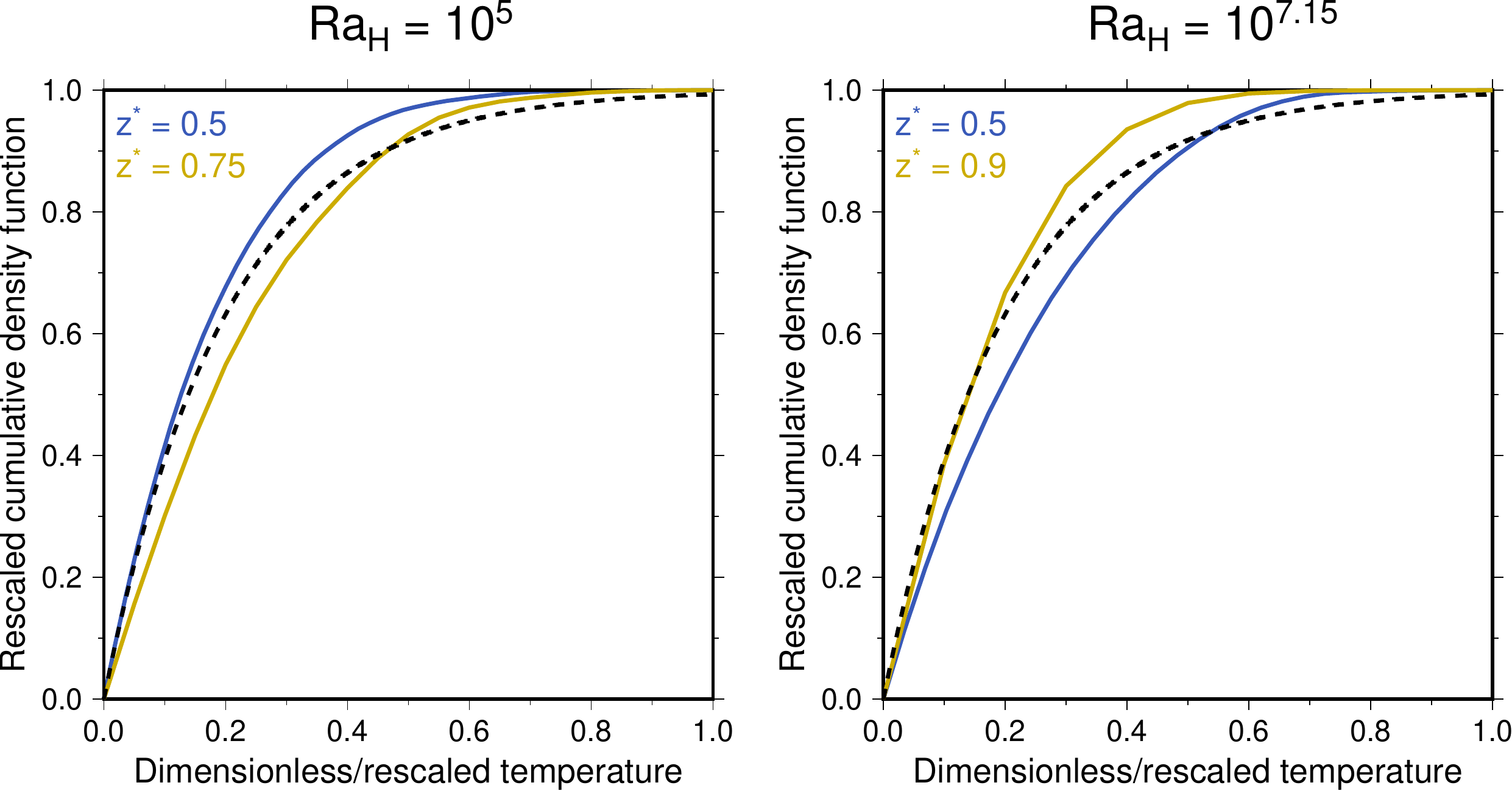}
\end{center}
\caption{\label{DistriT} Distributions of temperature from 3D numerical simulations taken at different depths for $Ra_{H} = 10^5$ (left panel) and $Ra_{H} = 10^{7.15}$ (right panel). The plots focus on the 5\% hottest temperatures, where melting is potentially occurring. When the distributions are rescaled, they show a similar behaviour that can be fitted (dashed line) with eq.~\ref{eqDistri}, here showed for $p = 5$.}
\end{figure}

To overcome these difficulties, we only focus on the hottest temperatures, here taken as the cdf between 0.95 and 1.
Interestingly, when these distributions are rescaled, i.e., the dimensions of both the x-axis and y-axis are modified in order to range from 0 to 1 (figure~\ref{DistriT}), they exhibit a similar trend, independently of the depth or $Ra_{H}$.
We can therefore estimate this part of the distribution using a simple function.
For instance, 
\begin{linenomath*} 
\begin{equation}
g^{*}_{\text{dist}} = 1 - \text{exp}(-p T^{*}_{rsc}),
\label{eqDistri}
\end{equation}
\end{linenomath*}
with $p$ a fitting parameter and $T^{*}_{rsc}$ the dimensionless and rescaled temperature, provides a good fit (figure~\ref{DistriT}).
The best-fit value of the parameter $p$ changes with the depth and $Ra_{H}$ considered and ranges typically between 1 and 10.
Here, for the sake of simplicity, we consider a constant value $p = 5$.
Note that we will discuss later the impact of $p$ on the melting distribution predicted by our approach.

\begin{figure}[t!]
\begin{center}
\includegraphics[width=0.9\textwidth]{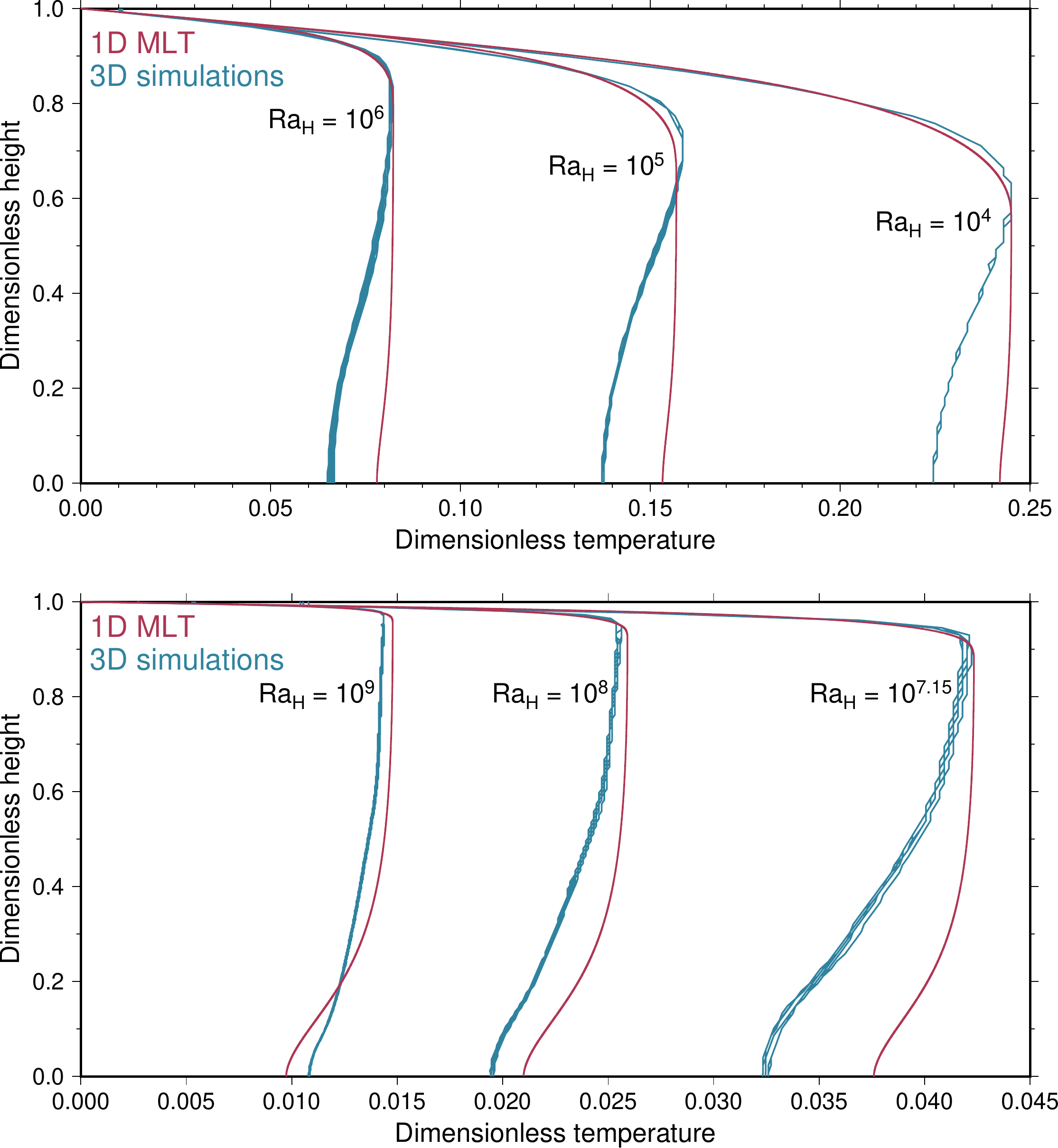}
\end{center}
\caption{\label{Profile_T95} Temperature profiles corresponding to a cumulative density function of 95\% from 3D numerical simulations (blue curves) and calculated by our approach (red curves) for various values of the Rayleigh-Roberts number ($Ra_{H}$). Profiles at different time-steps are reported to account for temporal variations. The red curves are calculated using eq.~\ref{eqT95} and the profiles predicted in figures~\ref{Profile_Tmoy} and~\ref{Profile_Tmax}.}
\end{figure}

To obtain the original distribution of temperature from the rescaled one, it is necessary to estimate the temperature profile corresponding to a cdf of 95\% ($T^{*}_{95}$).
Based on empirical observations (Supplementary Figure 2), we calculate this profile assuming, at first order, that
\begin{linenomath*} 
\begin{equation}
T^{*}_{95} = (1/3)\,T^{*}_{avg} + (2/3)\,T^{*}_{hot}.
\label{eqT95}
\end{equation}
\end{linenomath*}
Our predictions are plotted in figure~\ref{Profile_T95} along with the results from 3D numerical simulations.
For $Ra_{H}<10^{7}$, in the TBL, the profiles agree reasonably well with slight differences, equivalent to the ones found for the average (figure~\ref{Profile_Tmoy}) and hot (figure~\ref{Profile_Tmax}) temperature profiles.
For $Ra_{H}>10^{7}$, the agreement is less satisfactory with predictions that slightly overestimate the temperatures at the base of the TBL.
Nevertheless, close to the base of the TBL (where the differences are larger), the disagreement is lower than 5\% for all the cases studied.
Predictions given by eq.~\ref{eqT95} are therefore accurate enough for our purpose.

\subsection{Summary of the key stages of our method}

At that point, we have all the necessary ingredients to estimate, at given conditions, the distribution of the 5\% hottest temperatures and thus to estimate the generation of melt in a natural system.
For the sake of clarity, we provide in this section a brief summary of this process.
The first step is to build the dimensionless hot ($T^{*}_{hot}$) and horizontally averaged ($T^{*}_{avg}$) profile using 
\begin{linenomath*} 
\begin{equation}
\frac{\mathrm{d}T_{hot}^{*}}{\mathrm{d}z^{*}} - \frac{Ra_{H} l^{*}(z^{*})^{4}}{18}\biggl(\frac{\mathrm{d}T_{hot}^{*}}{\mathrm{d}z^{*}} - \frac{0.055}{z^{*\, 1.52}}\biggr)^{2} + 1.65 z^{*} = 0
\end{equation}
\end{linenomath*}
and
\begin{linenomath*} 
\begin{equation}
\frac{\mathrm{d}T_{avg}^{*}}{\mathrm{d}z^{*}} - \frac{Ra_{H} l^{*}(z^{*})^{4}}{18}\biggl(\frac{\mathrm{d}T_{avg}^{*}}{\mathrm{d}z^{*}} - \frac{0.05}{z^{*\, 1.55}}\biggr)^{2} + z^{*} = 0,
\end{equation}
\end{linenomath*}
respectively.
In these equations, the profile of the mixing length $l^{*}$ is indicated in figure~\ref{MLT_Kamata}, while its maximum value is given by eqs~\ref{eqLB_max} and~\ref{eqLB}, respectively.
Note that the parameter values are reported in Table~\ref{TabParam}.
The second step is to use eq~\ref{eqT95} to calculate $T^{*}_{95}$, the temperature profile corresponding to a cumulative density function of 95\%.
It is then possible to calculate at every depth the distribution of temperature between $T^{*}_{95}$ and $T^{*}_{hot}$ using eq~\ref{eqDistri}, with $p = 5$.
The last step is to dimensionalize the calculated distribution of temperature using the temperature scale of the system (eq~\ref{eqDTH}).
These distributions can then be used to estimate the generation of melt in a natural system.
However, our approach relies on a large number of assumptions, each one having intrinsic uncertainties.
It is therefore required to test the ability of our method to reproduce well the amount of melting that are potentially generated.

\section{Validation of our analytical approach}

The aim of this section is to validate the ability of our analytical approach to estimate the generation of melt.
In practice, this validation requires an application to a natural system.
Because our model assume a rather simple set-up, for instance, viscosity variations and compressibility effects are neglected, its application to an actual planetary body may not be fully appropriate.
To demonstrate its viability, we therefore decide to apply our analytical approach to a generic planetary mantle.
The application will be conducted in two stages.
First, we will compare 1D predictions and 3D numerical results at steady state for various $Ra_{H}$ in the range $10^{4}$--$10^{9}$.
This would allow us to confirm that, at first order, the 1D predictions are consistent with results of 3D simulations.
Then, we will use a more sophisticated model to estimate the evolution of a generic planetary mantle using both our analytical approach and the numerical code StagYY \cite{Tackley2008}. 

\subsection{Silicate mantle at steady state \label{secSteady}}
The first part of this application will be conducted as follows:
(i) we will calculate the analytical results corresponding to each of our 3D numerical simulations;
(ii) the analytical and numerical results will be applied to the generic mantle. In particular, we will estimate the proportion of material above the melting temperature as a function of depth;
(iii) we will compare the two sets of results to validate our analytical approach.
Using the line of reasoning developed in section~\ref{secMLT}, the calculation of the analytical results is straightforward.
It only requires to dimensionalize the results and set the solidus profile.

The temperature scale of the system is given in eq.~\ref{eqDTH} and depends on the heating rate ($H_{tp}$), the mantle depth ($d_{tp}$) and its thermal conductivity ($\lambda_{tp}$).
For the sake of example, we decide to study a 1000~km deep generic planetary mantle composed of silicates.
Under this assumption, typical values for the properties of silicates are reported in Table~\ref{TabProp} along with other important properties.
Note that different values could have been considered without affecting our conclusions.
The last parameter to calculate the temperature scale is $H_{tp}$.
A large range of values for $H_{tp}$ has been investigated.
From these results, we only report a few cases corresponding to the generation of a moderate amount of melt.
The quantity of melt generated is estimated by comparing the calculated temperatures to a solidus profile (Supplementary Figure 3) obtained from an interpolation of experimental works on terrestrial peridodites \cite{Hirschmann2000, Zhang1994, Andrault2011}, while the pressure is assumed to by hydrostatic.
As such, we do not, strictly speaking, estimate the amount of melt generated, but the proportion $X_{M}$ of material with a temperature larger than the solidus profile.
Note that the amount of melt generated will be investigated in the next section.

\begin{figure}[t!]
\begin{center}
\includegraphics[width=0.99\textwidth]{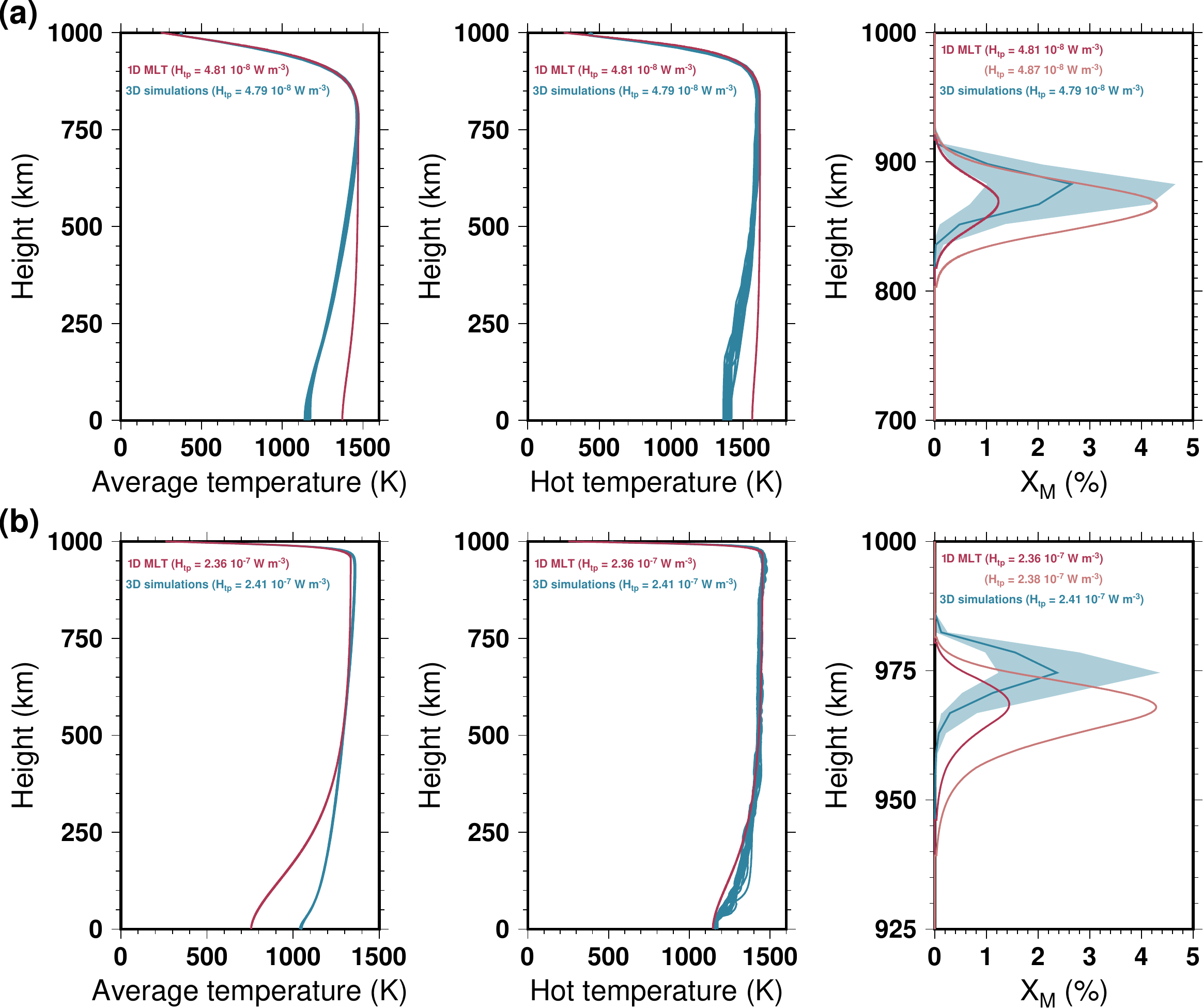}
\end{center}
\caption{\label{Profile_melt} Horizontally averaged (left column) and hot (central column) temperature profiles as well as $X_{M}$ the proportion of material whose temperature is larger than the solidus temperature (right column) for (a) $Ra_{H} = 10^{6}$ and (b) $Ra_{H} = 10^{9}$. Blue curves are obtained from 3D numerical simulations, while red curves are calculated with our analytical framework. In the right column, two different heating rates ($H_{tp}$) are considered in order to reproduce the temporal fluctuations exhibited by the 3D numerical results (blue shaded area), while the solid blue curve is the temporally averaged $X_{M}$.}
\end{figure}

The analysis of our results has shown that all the cases investigated could be separated into two groups depending on their $Ra_{H}$.
We therefore report in figure~\ref{Profile_melt} only one representative case for each group.
For cases with $Ra_{H}\leq 10^{7}$ (figure~\ref{Profile_melt}a), the predicted temperature profiles are in relatively good agreement with profiles obtained in numerical simulations.
As a result, the proportion of material with a temperature larger than the solidus profile agrees with the 3D numerical results.
To achieve this, however, we have considered a slightly different heating rate between the analytical and numerical results.
If we assume a single value, the general shape of the profiles is conserved, but the amplitude of $X_{M}$ is systematically underestimated or overestimated.
This is because most of the uncertainties in our analytical framework can be accounted for by changing the value of $H_{tp}$.
For instance, changing the value of $p$ in eq.~\ref{eqDistri} gives essentially the same results, except that it modifies the value of $H_{tp}$ required to obtain a good fit of the numerical results.
As a consequence, providing a scaling relationship for $p$, instead of assuming a constant value, may improve the fit at a given value of $H_{tp}$.
We however believe that it is unnecessary.
Assuming a constant value for $p$, we obtain for all the numerical simulations investigated an error on $H_{tp}$ lower than 4\% (an average of $\approx$2\%).
This error is totally negligible compared to the uncertainties on the appropriate heating rate for any planetary objects, even for Earth.
Furthermore, one may note that, when considering the decay of $^{238}$U, a 2\% error on $H_{tp}$ is roughly equivalent to a 100~Myrs temporal uncertainty.
For cases with $Ra_{H}\geq 10^{7}$ (figure~\ref{Profile_melt}b), the temperature profiles are also well reproduced by our analytical results.
Nevertheless, a slight disagreement exists between the two sets of results for $X_{M}$.
Indeed, the maximum of $X_{M}$ is slightly deeper in the analytical results (for $Ra_{H} = 10^{9}$, 32~km deep vs 25~km deep).
This difference is rather small (0.7\% of the mantle depth) and could be partly caused by the lower resolution of 3D simulations.
We therefore conclude that, despite the simplifications, our analytical framework is able to provide surprisingly good predictions for the location and proportion of material above the melting point.

\subsection{Evolution of a generic planetary mantle}

In the previous section, we have shown that our analytical approach should theoretically be able to provide a good estimate of the melt generation for a wide range of $Ra_{H}$.
We however used a simplified model assuming a steady-state without the effects of melting on the heat budget.
These results may thus in practice not be applicable to the evolution of planets.
Here, we therefore aim to verify the accuracy provided by our framework in the modelling of planets evolution.  
This will be done in several steps.
First, we develop an algorithm allowing to estimate the evolution of a planetary mantle including the effects of melting and secular cooling.
This model is then applied to the generic mantle described in the previous section.
As a second step, we conduct a numerical simulation of this generic mantle using StagYY \cite{Tackley2008}.
Finally, we compare the results obtained with the two approaches.
\subsubsection{Analytical estimation of the mantle evolution}

The evolution of our generic mantle is mainly governed by its conservation of energy,
\begin{linenomath*} 
\begin{equation}
\rho_{tp} C_{p, tp} \frac{\mathrm{d}T}{\mathrm{d}t} = -\frac{\mathrm{d}F_{cond}}{\mathrm{d}z} - \frac{\mathrm{d}F_{conv}}{\mathrm{d}z} + H_{rad} - \frac{P_{melt}}{V},
\label{eqAnal}
\end{equation}
\end{linenomath*}
with $H_{rad}$ the heating rate produced by radioactive heating, $P_{melt}$ the power used for melting, $V$ the volume of the mantle, and $t$ time.
This equation \textit{is a priori} different from the equation used in our analytical framework (eq~\ref{eqEnergy}), so that a direct application may not be possible.
However, one may note that if we simply write,
\begin{linenomath*} 
\begin{equation}
 H_{melt} = \frac{P_{melt}}{V},\hspace{0.3cm}H_{sec} = \rho_{tp} C_{p, tp} \frac{\mathrm{d}T}{\mathrm{d}t},\hspace{0.3cm}\text{and}\hspace{0.3cm}H_{eff} = H_{rad} -  H_{melt} - H_{sec},
\label{eqHeff}
\end{equation}
\end{linenomath*} 
then eq~\ref{eqAnal} implies
\begin{linenomath*} 
\begin{equation}
0 = -\frac{\mathrm{d}F_{cond}}{\mathrm{d}z} - \frac{\mathrm{d}F_{conv}}{\mathrm{d}z} + H_{eff},
\label{eqAnal_F}
\end{equation}
\end{linenomath*}
which is identical to eq~\ref{eqEnergy}.
It is therefore possible to use our framework to estimate the distribution of temperature, provided that the effective heating rate $H_{eff}$ is used.

The effective heating rate is composed of three components (eq.~\ref{eqHeff}).
The first component, the radioactive heating contribution, is calculated using
\begin{linenomath*} 
\begin{equation}
H_{rad} = H_{tp0}\, \text{exp}\biggl(-\text{log}(2) \frac{t}{t_{1/2}}\biggr),
\label{eqHrad}
\end{equation}
\end{linenomath*}  
with $H_{tp0}$ the initial heating rate and $t_{1/2}$ the half-life of the radioactive isotopes.
For the sake of simplicity, we only consider the presence of $^{238}$U and its initial abundance is set to obtain a moderate amount of melting.
The second component corresponds to the power associated with the melt generation,
\begin{linenomath*} 
\begin{equation}
H_{melt} = \frac{L_{tp} \rho_{tp} \Delta V_{melt}}{\Delta t\, V},
\label{eqHmelt}
\end{equation}
\end{linenomath*}
with $\Delta V_{melt}$ the volume of melt generated during $\Delta t$, $L_{tp}$ the latent heat of fusion, and $\Delta t$ the time-step of the evolution model.
Note that we here assume that all the material above the solidus is fully molten and instantaneously extracted to the surface (section~\ref{secLimitMelting} provides a detailed discussion on this assumption).
Therefore, the determination of $\Delta V_{melt}$ only requires to compare the solidus profile with the estimated distribution of temperature.
The third component corresponds to secular cooling.
In practice, we calculate this contribution using,
\begin{linenomath*} 
\begin{equation}
H_{sec} = \rho_{tp} C_{p, tp} \frac{\Delta T_{vol}}{\Delta t},
\label{eqHsec}
\end{equation}
\end{linenomath*}
with $\Delta T_{vol}$ the variation of the volume average temperature between the time $t$ and $t + \Delta t$.
Note that all the properties of the model is reported in Table~\ref{TabProp}.

Following eqs.\ref{eqHmelt} and \ref{eqHsec}, the effective heating rate $H_{eff}$ is depending on the thermal state of the mantle, while $H_{eff}$ is required to estimate the distribution of temperature.
One has therefore to follow an iterative process to obtain the actual thermal state (so $H_{eff}$) of the system.
In order to improve the speed and accuracy of our numerical scheme, we decided to calculate \textit{a priori} the temperature and melting profiles for a large range of $H_{eff}$.
As a result, at each time-step, the actual $H_{eff}$ can simply be obtained by solving eq.~\ref{eqAnal_F} using the pre-calculated data.
Once the value of $H_{eff}$ is obtained, we move to the time $t + \Delta t$ and repeat the process until the full evolution is calculated.

An interesting feature of this model is that the effective heating rate decreases, at first order, with $\Delta t$ (eqs.~\ref{eqHmelt} and \ref{eqHsec}).
It is therefore possible to lower the amount of melt produced by reducing the time-step used.
This is a crucial feature because our framework is only able to predict, at a given depth, a maximum melt proportion of 5\%.
Decreasing the time-step can therefore be used to overcome this limitation (see section~\ref{secLimitMelting} for more details).
A drawback, however, is that the required computational time increasing with decreasing value of the time-step.
We therefore use a short time-step in eq.~\ref{eqHmelt}, set by the need to have a low proportion of melting, while we use a time-step $\Delta t_{ev} = 10 000$~years for the time-stepping procedure (and in eq.~\ref{eqHmelt}).
This should have no impact on the results as long as the change of $H_{eff}$ remains modest between $t$ and $t+\Delta t_{ev}$.

\subsubsection{Numerical simulations of the mantle evolution}

The numerical simulation is designed to reproduce as closely as possible our analytical framework.
As such, we consider a simplified version of melting, where all the volcanism is extrusive, all material above solidus is fully molten, and melt is instantaneously extracted to the surface.
The rock properties are given in Table~\ref{TabProp} and the heating rate is specified in eq.~\ref{eqHrad}.
The initial temperature condition is obtained by running a numerical simulation, with a constant heating rate $H_{tp0}$, until a steady-state is reached.
This has the advantage to decrease the transient effects at the start of the simulation (see section~\ref{secLimitTransient} for more details).
The grid resolution and aspect ratio are 512$\times$256$\times$256 and 4:2, respectively.
Grid refinement is used at the surface to improve the resolution of the thermal boundary layer.

\subsubsection{Comparison between 1D analytical evolution and 3D simulation}
The comparison between the two approaches can be done using a variety of characteristics.
Here, we have selected the following properties: (i) the maximum and volume average temperature in the mantle, to asses the thermal evolution; (ii) the depths where melting occurs and the rate of melt production $R_{melt}$, to assess the melt generation.
These properties are straightforward to obtain except the melting rate that is calculated at each time-step using
\begin{linenomath*} 
\begin{equation}
R_{melt} = \frac{\rho_{tp} C_{p, tp}}{\Delta t V L_{tp}} \int_{V} (T - T_{sol})  dV,
\label{eqErupta}
\end{equation}
\end{linenomath*}
with $T$ the local temperature, and $T_{sol}$ the local solidus temperature.

We report in figure~\ref{Comparison} the evolution of the four selected properties for both the 1D analytical model and the 3D simulation.
We first note that the evolution of the maximum temperature is fairly well reproduced by the analytical approach with a typical deviation lower than 50~K ($\approx$6\%).
By contrast, the volume average temperature is systematically underestimated by $\sim$75~K ($\approx$10\%).
The latter is however not surprising as our analytical approach tends to underestimate the temperature in the convective interior for very large $Ra_{H}$ (figure~\ref{Profile_Tmoy}).
A close inspection of these plots reveals two important shortcomings of our analytical approach:
(i) the analytical approach predicts an almost constant temperature and melting depths for almost 1~Grys, whereas it is continuously decreasing in the simulation. 
This difference may be due to the difficulty of accounting for the dynamic nature of a convective system in an analytical approach based on a sequence of static snapshots (see section~\ref{secLimitMelting} for more details).
(ii) even though we tried to reduce the transient effects, we can still observe a slight transient phase in the first 200~Myrs.
Transient effects may be important for the early evolution of planetary bodies, while being very difficult to account for in 1D analytical models (see section~\ref{secLimitTransient} for more details).
With these two shortcomings in mind, we can now describe the evolution of the melt generation.
In both the analytical approach and 3D simulation, melting is present for an extended period of time and stops at $\sim$1.6~Grys.
The depths where melting occurs is well reproduced by the analytical approach, especially the initial depth range where melting occurs ($\sim$60-100~km) and the decreased depth range with time.
The only notable difference concerns the amplitude of the melting rate that is systematically underestimated by a factor 3 in the analytical calculation.

The comparison between the 1D analytical evolution and 3D simulation has shown that some properties, such as the maximum temperature or the depth range where melting occurs, are relatively well reproduced by our analytical approach, while other properties, such as the volume average temperature or total amount of melting, are subject to clear deviations.
Note however that the loss in accuracy is compensated by a large gain in speed and simplicity.  
Our analytical approach stands therefore as a powerful tool to study the evolution of planetary bodies.

\section{Discussion \label{secDiscu}}
\subsection{Limitations \label{secLimit}}
The approach developed in this study is subject to several important limitations.
While some of these limitations can be easily alleviated, some others are inherent to the model and should be bear in mind.
We provide in this section a brief discussion of these limitations.
\subsubsection{Applicability domain \label{secLimitCondition}}
To start the discussion, it is important to mention that the laws built in this work are only valid for the specific conditions investigated.
In particular, it would be hazardous to apply our results to systems with $Ra_{H}$ much larger than 10$^{9}$.
Indeed, the calculated temperature profiles are very sensitive to the value of $\delta_{TBL}$.
So the inevitable error on $\delta_{TBL}$ made when extrapolating the scaling laws may lead to very large errors on the calculated temperature profiles. 
\subsubsection{Transient effects \label{secLimitTransient}}
Transient effects are particularly important in the early evolution of planets.
Right after the planet formation, the whole planet, including the mantle, is unlikely to be at steady state, but rather in a much hotter state.
The planetary mantle may even be fully molten and experience a magma ocean event \cite<e.g.,>[]{Abe1993}.
As a consequence, the mantle evolution is characterized by a first phase, possibly lasting for several hundred million years, where it cools down until reaching a steady state.
This first phase may be crucial in planetary evolution because it can potentially impact the whole subsequent evolution, as it has been postulated for instance for the Moon \cite{Laneuville2013} or Earth \cite{Ballmer2017}.
However, it is currently not possible to account for these transient effects in 1D analytical models.
This would require additional studies specially devoted to investigate transient convection, which are, to our knowledge, not available.
In particular, it would be interesting to check whether scaling laws established at steady state remain valid in the transient phase.
Nevertheless, one may keep in mind that we have very few constraints on the early evolution of planets.
Therefore, even if we could model accurately these transient effects, a lot of uncertainties would remain on this early evolution because of the lack of constraints.
\subsubsection{Secular cooling \label{secLimitSec}}
Results of our 1D analytical evolution show a systematic underestimation of the volume average temperature ($\sim$75~K).
Although it does not directly impact the generation of melting, it still influences the results as it may change the importance of secular cooling (eq.~\ref{eqHsec}).
In our case, the temporal derivative of the volume average temperature, which set the amplitude of secular cooling, is similar in the 1D analytical evolution and 3D numerical simulation (figure~\ref{Comparison}b).
Thus, secular cooling should be correctly estimated.
This may however not be always true.
In particular, when the Rayleigh-Roberts number is changing importantly throughout the calculated evolution, one may expect a non-systematic error on the volume average temperature, which in turn may lead to a significant error on the amplitude of secular cooling.
Nevertheless, for a typical terrestrial planet, we do not expect large temporal variations of the Rayleigh-Roberts number so that our approach should be accurate.
One should however remain careful when investigating evolutions exhibiting large temporal variations of the Rayleigh-Roberts number.
\subsubsection{Melting process \label{secLimitMelting}}
Our model assumes a simple version of melting where at each time-step all the melt is reaching instantaneously the surface (extrusive volcanism).
This assumption underlies several important simplifications.
First, in a planetary setting, a non-negligible proportion of the melt would actually not reach the surface but would remain in the mantle and solidify (intrusive volcanism).
The exact ratio between intrusive and extrusive volcanism is difficult to estimate, such that it is often considered as an input parameter in 3D simulations.
Second, melting ascent is actually not instantaneous, although it is much faster than the typical convective time-scale.
Last, all material above the solidus is not fully molten but experiences partial melting producing chemical differentiation.
One may note that these limitations are not peculiar to our analytical approach as they are also present in 3D numerical simulations.
Different methods have been developed to account for these complexities that could be included in our model.
Here, the use of a simple version of melting was not dictated by an intrinsic limitation of our model, but because the aim of this study is only to show the validity of our approach and not to provide a realistic modelling of mantle evolution.

Our analytical approach has however an additional intrinsic limitation that induces a connection between the time-step used and the melt generation.
A simple way to identify this connection is by considering the extreme case of a very small time-step $\Delta t$.
For such a case, following eqs.~\ref{eqHeff} and~\ref{eqHmelt}, the volume of melt produced $\Delta V_{melt}$ has to be very low in order to maintain a reasonable value for the effective heating rate  $H_{eff}$.
In this extreme case, the maximum amount of melting would therefore be low and the melt generation would only occur at a specific depth.
This limitation comes from the dynamic nature of a convective system and can be understood by considering the evolution of a hot upwelling.
In a first stage, the hot upwelling is rising until to cross the solidus and to start generating melt.
However, the first generation of melt would not stop its motion, the hot upwelling would therefore keep rising as long as it is buoyant and produces melt over a certain depth range.  
Unfortunately, this dynamic process cannot be reproduced in an analytical approach relying on static snapshots of the convective system.
Here, we thus tried to account for this limitation by a fine tuning of the time-step value.
More specifically, our algorithm is designed to preferentially select a time-step value leading to a maximum amount of melt around 4.5\%.
Nevertheless, a different choice of value could be made inducing significant changes in the calculated evolution.

\begin{figure}[t!]
\begin{center}
\includegraphics[width=0.9\textwidth]{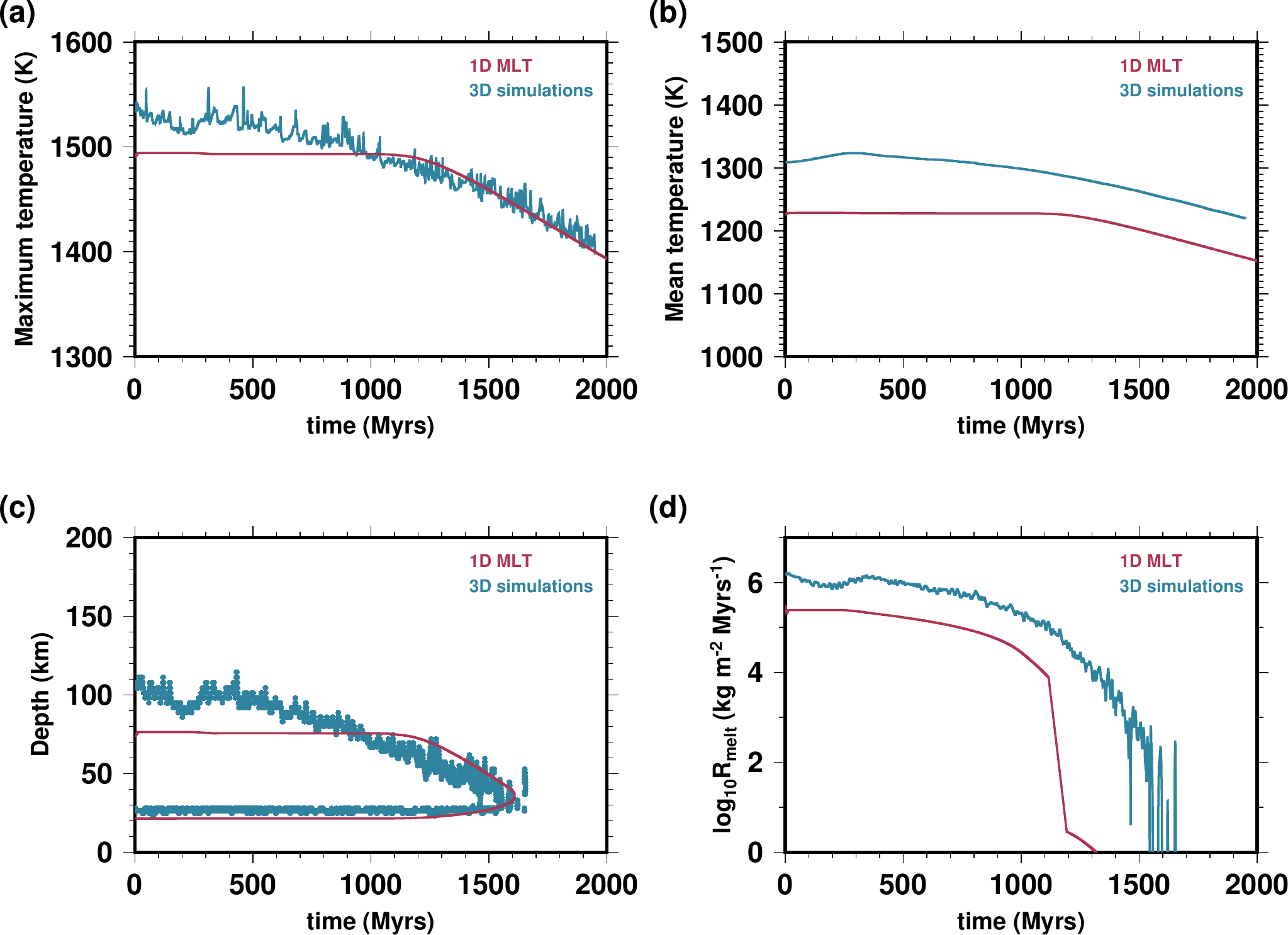}
\end{center}
\caption{\label{Comparison} Evolution of the (a) maximum temperature, (b) the volume average temperature, (c) the depths where melting occurs, and (d) the melting rate for the generic mantle described in section~\ref{secSteady}. Blue curves are obtained from a 3D numerical simulation, while red curves are calculated with our analytical framework.}
\end{figure}

\begin{figure}[t!]
\begin{center}
\includegraphics[width=0.9\textwidth]{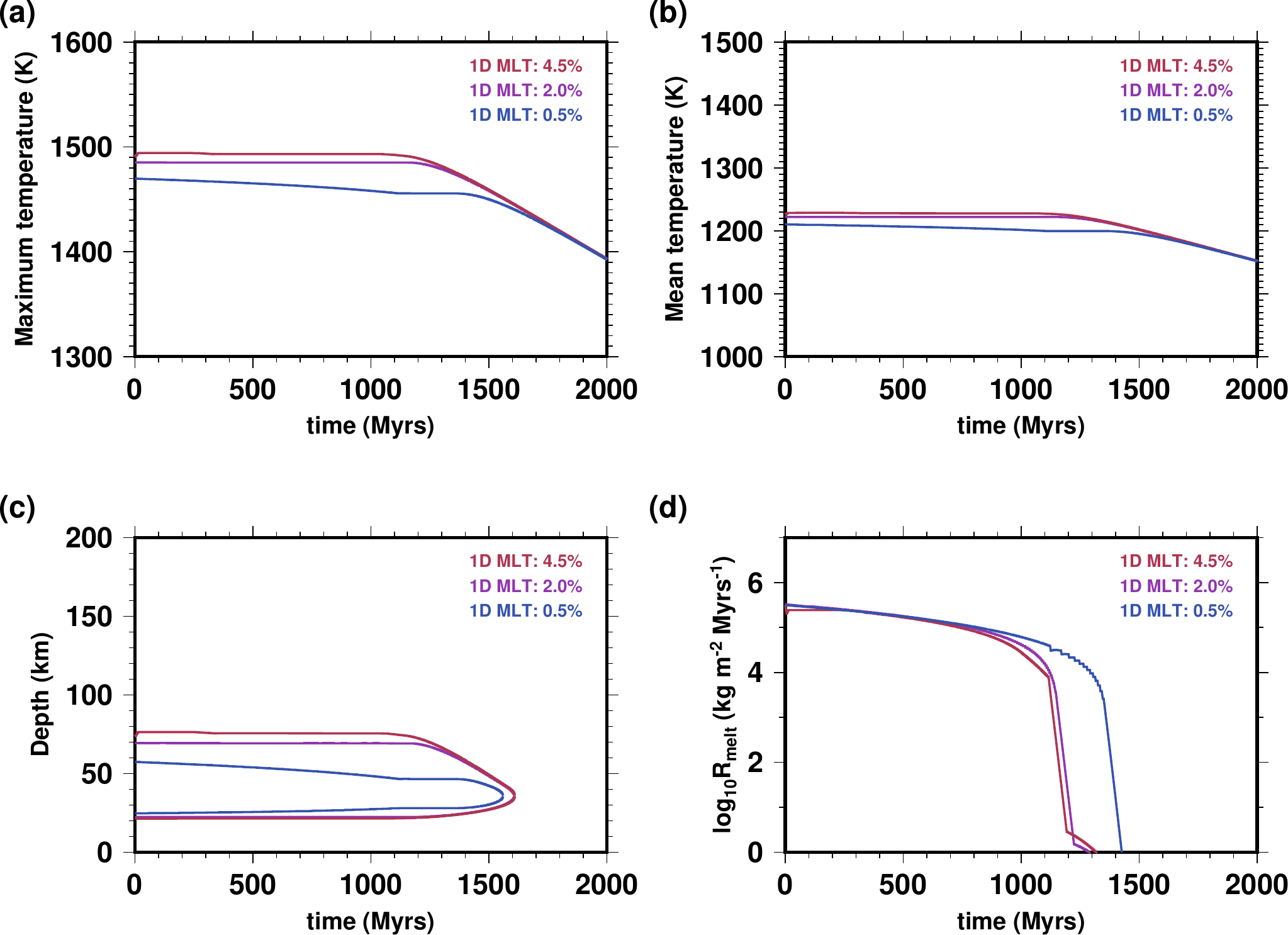}
\end{center}
\caption{\label{UncertaintyAmp} Evolution of the (a) maximum temperature, (b) the volume average temperature, (c) the depths where melting occurs, and (d) the melting rate estimated using our analytical approach for the generic mantle described in section~\ref{secSteady}. The red curve is from figure~\ref{Comparison}, while the purple and blue curves are obtained favouring a maximum amount of melting of 2.0\% and 0.5\%, respectively.}
\end{figure}

In order to evaluate the consequences of this choice on our model results, we plotted in figure~\ref{UncertaintyAmp} the calculated evolutions assuming a maximum amount of melt of 4.5\% (as in figure~\ref{Comparison}), 2.0\% and 0.5\%.
As predicted, the depths where melting occurs is the most impacted property with a reduction of the depth range as the maximum amount of melting allowed decreases.
Interestingly, this reduction is mainly caused by the decrease of the estimated maximum depth of melting for a low maximum amount of melt ($<$2\%).
By contrast, the estimated minimum depth of melting is only slightly affected indicating that it is a robust output of our model.
We also note that the maximum temperature follows a similar behaviour than the maximum depth of melting, while the volume of melt produced and the duration of the melting event are hardly affected.

Overall, we conclude that only the maximum depth where melting occurs and the maximum temperature may be significantly affected by this limitation.
Nevertheless, our choice for the maximum amount of melt allowed (i.e., 4.5\%) seems to be high enough to provide a reasonable estimate for these properties, as shown by the results in figure~\ref{Comparison}.
We however note that it remains unclear whether this would still be valid when applied to a different system.
One should therefore be cautious when analysing the maximum depth where melting occurs.

\subsection{Implications for melting in exoplanets}

A unique characteristic of our approach is to provide a fast and precise estimate of the melt generation.
This ability may be particularly useful when studying poorly constrained terrestrial planets, since a thorough numerical investigation of such planets would require a large calculation power.
Here, we propose to show the strength of our approach by investigating the generation of melting in exoplanets.
A previous work \cite{Vilella2017} has already predicted the occurrence of melting as a function of the planetary radius and heating rate.
Combining this work with our approach, we further predict the melt volume and melting depths.

Following the model used by \citeA{Vilella2017}, we consider a differentiated terrestrial planet with a radius $R_{tp}$ and a heating rate $H_{tp}$.
The mass of the planet ($M_{tp}$) is obtained using the mass-radius relationship obtained by \citeA{Valencia2006} for Earth-like planets, 
\begin{linenomath*} 
\begin{equation}
\frac{R_{tp}}{R_T} = \biggl(\frac{M_{tp}}{M_{T}}\biggr)^{0.27},
\end{equation}
\end{linenomath*}
where $R_T$ and $M_T$ are the radius and mass of the Earth, respectively.
This allows us to calculate the acceleration of gravity $g_{tp} = G\, M_{tp}/ R_{tp}^{2}$, with $G$ the gravitational constant.
The mantle depth ($d_{tp}$) is estimated using \cite{Valencia2006},
\begin{linenomath*} 
\begin{equation}
\label{coreValencia}
d_{tp} = R_{tp} - R_{c,tp} \hspace{0.5cm} \text{with} \hspace{0.5cm}R_{c,tp} = 3.5\, 10^3\, \biggl(\frac{R_{tp}}{R_{T}}\biggr)^{0.926},
\end{equation}
\end{linenomath*}
the core radius (all the distances are in kilometres).
By contrast with \citeA{Vilella2017}, we do not consider a temperature dependent viscosity nor the effects of spherical geometry.
Furthermore, values for rock properties are listed in Table~\ref{TabProp} and slightly differ from the ones used in \citeA{Vilella2017}, in particular the pressure is here assumed to be hydrostatic.

\begin{figure}[t!]
\begin{center}
\includegraphics[width=0.9\textwidth]{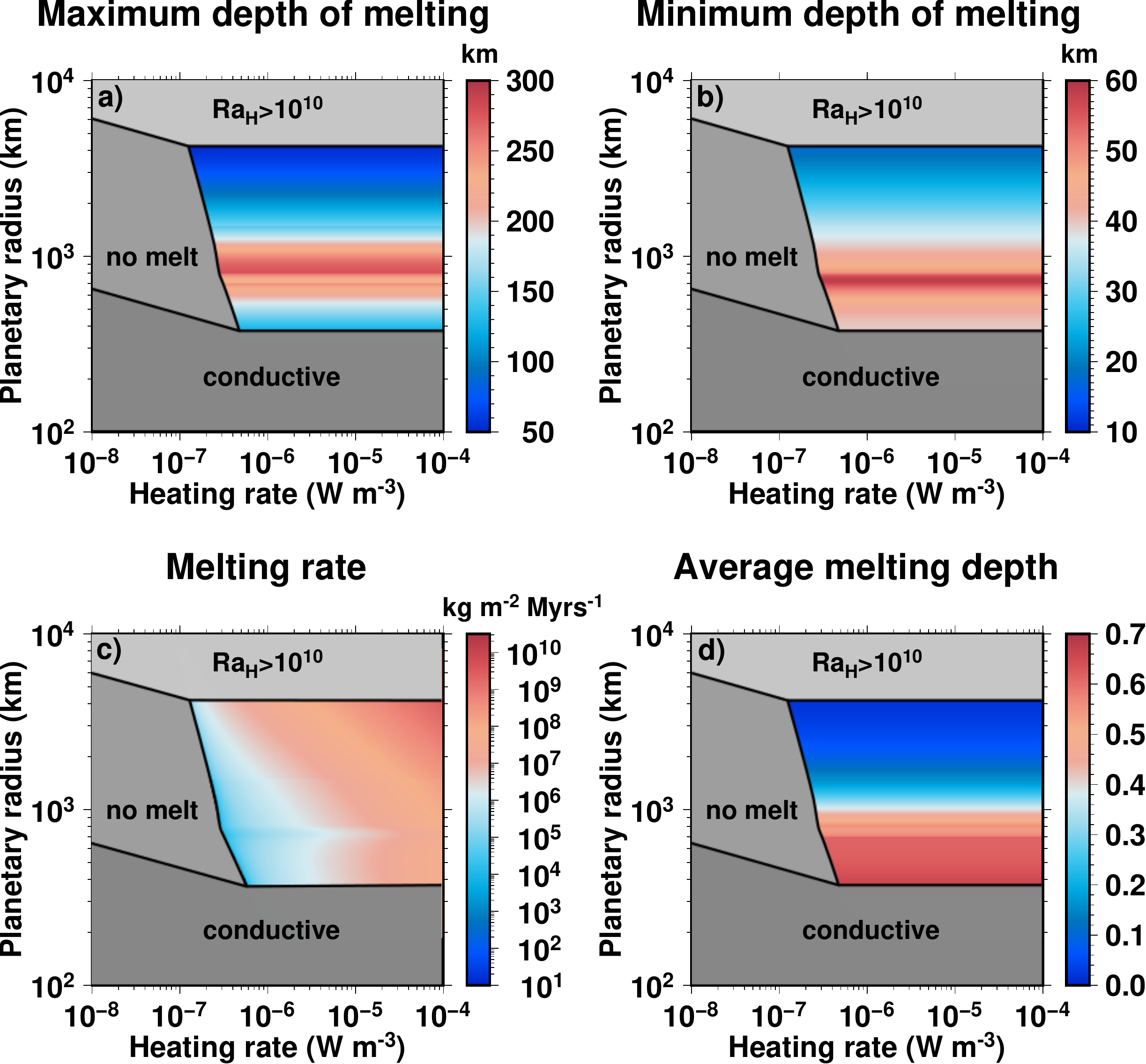}
\end{center}
\caption{\label{MeltingDiagram} Diagram showing the estimated (a) maximum depth of melting, (b) minimum depth of melting, (c) melting rate, and (d) dimensionless average melting depth as a function of the planetary radius and heating rate. We exclude the cases where the natural system is either conductive ($Ra_{H}<868$), or has a Rayleigh-Roberts number so high that our approach cannot be applied ($Ra_{H}>10^{10}$). The details of the modelling and parameters values are given in the text.}
\end{figure}

Using this set of parameters combined with our analytical approach, we estimate the generation of melting for a large range of planetary radius and heating rates assuming a maximum amount of melt of 4.5\% (see section~\ref{secLimitMelting}).
As a first step, we identify the conditions for which the Rayleigh-Roberts number is either too high to apply our approach (here chosen to be $Ra_{H}>10^{10}$), or too low to allow for convection ($Ra_{H}<868$).
As a second step, we exclude all the cases where no melt is generated to focus only on the conditions for which melting is present (figure~\ref{MeltingDiagram}).
In that regime, a surprising result is the absence of variations in the melting depths with increasing heating rate, while one may have expected a widening.
This is actually caused by the simplified version of melting considered here (see section~\ref{secLimitMelting}).
Furthermore, for heating rates as large as $10^{-4}$~W~m$^{-3}$, one may expect the appearance of a local magma ocean.
As a matter of fact, considering $R_{tp} = 1000$~km and $H_{tp} = 10^{-4}$~W~m$^{-3}$, a volume equivalent to the whole mantle should melt every 10~Myrs.
This may be difficult to reconcile with solid-state convection.
Our results can therefore be seen as a preliminary step that require to be further combined with a more sophisticated model of melting in order to reach a more precise assessment of melting in exoplanets.

Nevertheless, the present results already provide crucial and robust information that deserved to be discussed.
For instance, the depth range where melting occurs exhibit two different trends with increasing planetary radius.
For $R_{tp}<1000$~km, the depth range becomes deeper with increasing planetary radius.
In that case, the solidus temperature varies only slightly with depth due to the low acceleration of gravity.
As a result, the melting occurs from the base of the mantle to the base of the thermal boundary layer.
In other words, the maximum depth of melting is simply the mantle depth, while the minimum depth of melting varies as the thickness of the thermal boundary layer \cite<which is increasing with the planetary radius following>[]{Vilella2017}.
One way to verify this observation is with the dimensionless average melting depth (figure~\ref{MeltingDiagram}d) that remains constantly high ($>$0.6) in that regime.
For $R_{tp}>1000$~km, the depth range becomes shallower with increasing planetary radius.
This is because the solidus temperature is starting to increase with the planetary radius.
Melting therefore requires a higher effective heating rate, inducing a larger $Ra_{H}$, so a thinner thermal boundary layer.
This trend is likely to remain valid for $Ra_{H}$ larger than $10^{10}$ as long as the trend of the solidus profile with pressure is not significantly modified, which is unlikely to happen even for large exoplanets \cite{Stixrude2014}.

\section{Conclusion}

We have developed a new analytical approach, partly based on an extended version of the mixing length theory, able to estimate the distribution of the hottest temperatures for a purely volumetrically heated system.
These temperature distributions have then been compared to results obtained from 3D numerical simulations for a large range of $Ra_{H}$.
The final objective of this study was to assess the melt generation in 1D analytical models.
We have therefore proposed a 1D analytical approach that accounts for the effects of melting, secular cooling and radioactive heating. 
To demonstrate its relevance, we applied this model to a generic planetary mantle and conducted a 3D simulation mimicking these conditions.
The comparison between the analytical results and the 3D simulation are very encouraging.
In particular, the evolution of the maximum temperature and the depth range where melting occurs is well reproduced by our analytical approach.
Note however that the volume average temperature is systematically underestimated by $\sim$100~K and that the total amount of melting was 3 times lower than in the 3D simulation.
Despite these slight disagreements, our analytical approach stands as a powerful tool to study planetary evolution:
(i) it is the only analytical approach able to constrain a part of the temperature distribution, i.e., that is able to model melt generation properly;
(ii) compared to 2D or 3D simulations, 1D analytical models are much faster so that a much larger parameter space can be explored;
(iii) our analytical approach can be combined with a sophisticated modelling of melt generation in order to investigate the chemical evolution of planetary mantles.
The latter point is crucial for our understanding of planetary evolution, while it seems out of reach for current 2D and 3D simulations.
This study, however, considers a purely volumetrically heated fluid, which may not be appropriate to model accurately a planetary body.
Additional work is therefore required to extend our analytical approach to a system appropriately depicting a planetary mantle or an icy shell. 
\clearpage
\newpage
\begin{table}
{ \small
\begin{tabular}{l c c c c}
                    &                    & $Ra_{H}<10^{5}$           & $10^{5}\leq Ra_{H}<10^{7}$ & $10^{7}\leq Ra_{H}$ \\
                    &                    &                           &                            & \\
Average            & $\delta_{TBL}^{*}$ & 0.8241$Ra_{H}^{-0.06637}$ & 4.4412$Ra_{H}^{-0.2203}$   & 5.6885$Ra_{H}^{-0.2352}$\\
temperature profile & $c$                & 0.05                      & 0.05                       & 0.05\\
                    & $d$                & 1.55                      & 1.55                       & 1.55\\
                    &                    &                           &                            & \\
Hot                 & $\delta_{TBL}^{*}$ & 1.7881$Ra_{H}^{-0.1558}$  & 4.3793$Ra_{H}^{-0.234}$    & 4.6238$Ra_{H}^{-0.2361}$\\
temperature profile & $c$                & 0.055                     & 0.055                      & 0.055\\
                    & $d$                & 1.52                      & 1.52                       & 1.52\\
\end{tabular}\\ }
\caption{\label{TabParam} Parameters of the modified mixing length theory (see text for more details).}
\end{table}
\clearpage
\newpage
\begin{table}
\begin{tabular}{c c c c}
Parameter                     & Symbol         & Value              & Unit\\
                              &                &                       & \\
Mantle depth                  & $d_{tp}$       & 1000                  & km\\
Gravity acceleration          & $g_{tp}$       & 3.0                   & m~s$^{-2}$\\
Surface temperature           & $T_{surf, tp}$ & 250                   & K\\
Density                       & $\rho_{tp}$    & 3400                  & kg~m$^{-3}$\\
Thermal conductivity          & $\lambda_{tp}$ & 3.0                   & W~m$^{-1}$~K$^{-1}$\\
Thermal expansion coefficient & $\alpha_{tp}$  & 5$\times$10$^{-5}$    & K$^{-1}$\\
Thermal diffusivity           & $\kappa_{tp}$  & 7$\times$10$^{-7}$    & m$^{2}$~s$^{-1}$\\
Specific heat capacity        & $C_{p, tp}$    & 1260                  & J~kg$^{-1}$~K$^{-1}$\\
Viscosity                     & $\eta_{tp}$    & 10$^{20}$             & Pa~s\\
Latent heat of fusion         & $L_{tp}$       & 600                   & kJ~kg$^{-1}$\\
Half-life $^{238}$U           & $t_{1/2}$      & 4.468                 & Gyrs\\
Initial heating rate          & $H_{tp0}$      & 2.45$\times$10$^{-7}$ & W~m$^{-3}$
\end{tabular}
\caption{\label{TabProp} Model parameters used for the generic planetary mantle studied.}
\end{table}
\clearpage
\newpage
\clearpage
\newpage
\clearpage
\newpage
\clearpage
\newpage
\clearpage
\newpage
\clearpage
\newpage
\clearpage
\newpage
\clearpage
\newpage
\clearpage
\newpage
\clearpage
\newpage
\clearpage
\newpage
\clearpage
\newpage
\appendix 
\section{Solving the governing equation}
The resolution of eq.~\ref{eqMLT} is achieved using a finite difference approach.
More specifically, the discretization of eq.~\ref{eqMLT} implies,
\begin{linenomath*} 
\begin{equation}
\frac{T_{i+1}^{*} - T_{i}^{*}}{dz} - \frac{Ra_{H} l^{*\,4}_{i}}{18}\biggl(\frac{T_{i+1}^{*} - T_{i}^{*}}{dz} - f^{*}(z^{*}_{i}) \biggr)^{2} + z^{*}_{i} = 0,
\end{equation}
\end{linenomath*}
where $T_{i}^{*}$ and $l^{*}_{i}$ are the temperature and mixing length at height $z^{*}_{i}$, and $T_{i+1}^{*}$ the temperature at height $z^{*}_{i+1} = z^{*}_{i} + dz$, with $dz$ a small increment.
This equation allows to calculate $T_{i}^{*}$ from $T_{i+1}^{*}$ by solving a second order polynomial equation.
A difficulty, however, is to choose the correct solution from the two solutions of the polynomial equation.
To do so, we select the solution given the best continuity in term of temperature gradient. 
As a result, the increment $dz$ has to be very small in order to ensure that the correct solution is chosen.
For reference, we typically use 50000 vertical points to safely prevent any possible problems.
Note that the same method can be followed to calculate the hot temperature profile from eq.~\ref{eqMLT_max}.
\section*{Acknowledgements}
This research was funded by JSPS KAKENHI Grant Number JP19F19023.
The data used for generating the figures are available for academic purposes \cite{Vilella2020d}.
All required details to reproduce the analytical results are given in the text, while the numerical results have been detailed in a previous publication \cite{Vilella2018b}.
Note that the program used to calculate the analytical evolution reported in Figure 10 is available for academic purposes \cite{Vilella2020d}.
The numerical code used to obtain the numerical results is not publicly available but was thoroughly described in \citeA{Tackley2008}.
\bibliography{/home/kenny/Documents/Work/Article/biblio}
\end{document}